\documentclass[a4paper,11pt]{article}
\usepackage{geometry}
\usepackage{makeidx}
\usepackage[normalem]{ulem}
\usepackage{lineno}
\usepackage[dvips]{graphicx}
\usepackage{color}
\usepackage[numbers,sort&compress,comma]{natbib}
\usepackage{amssymb,amsmath,fullpage,graphicx,pstricks,sectsty,subfigure}
\usepackage{fancyvrb}

\usepackage{amsfonts}
\usepackage{mathtools}
\usepackage{mathptmx}
\usepackage{euler}

\def\eq{\begin{equation}}
\def\en{\end{equation}}

\usepackage{amsmath}

\title{
A {generalized} Mittag-Leffer function to describe nonexponential chemical effects}

\author{
Nelson H. T. Lemes$^{\rm a}$\footnote{\texttt{nelson.lemes@unifal-mg.edu.br}},
Jos\'e Paulo C. dos Santos$^{\rm b}$ and Jo\~ao P. Braga$^{\rm c}$\\
  \small{$^{\rm a}$Instituto de Qu\'{\i}mica,}
  \small{Universidade Federal de Alfenas - UNIFAL}\\
  \small{Brazil, Alfenas, MG (37130-000)}\\
  \small{$^{\rm b}$Instituto de Ci\^encias Exatas,}
  \small{Universidade Federal de Alfenas - UNIFAL}\\
  \small{Brazil, Alfenas, MG  (37130-000)}\\
  \small{$^{\rm c}$Departamento de Qu\'{\i}mica,}
  \small{Universidade Federal de Minas Gerais - UFMG}\\
  \small{Brazil, Belo Horizonte, MG (31270-010)}
}
\date{}

\begin{document}
\maketitle
\clearpage
%\tableofcontents
%\newpage

%\begin{linenumbers}
%%%%%%%%%%%%%%%%%%%%%%%%%%%%%%%%%%%%%%%%%%%%%%%%%%%%%%%%%%%%%%%
%\subsection{Abstract}
\begin{center}{\bf Abstract}\end{center}
%contextualização e lacuna%%%%%%%%%%%%%%%%%%%%%%%%%%%%%%%%%%%%%%%%%%%%%%%%%%%%%%%%%%%%%%%%%%%%%%%%%%%%%%%%%%%%%%%
%\textcolor{red}{CONTEXTUALIZAR} The chemical luminescence decay was always assumed to be exponential, \textcolor{red}{LACUNA} {\bf but} recently it was 
%found experimental data that shows a different behavior at longer time.
%for a long time enough, a different behavior. %to observe nonexponential behavior
%proposta%%%%%%%%%%%%%%%%%%%%%%%%%%%%%%%%%%%%%%%%%%%%%%%%%%%%%%%%%%%%%%%%%%%%%%%%%%%%%%%%%%%%%%%%%%%%%%%%%%%%%
%\textcolor{red}{PROPOSTA} 
In this paper a differential equation with noninteger order was used to model an anomalous luminescence decay process. Although this process is in principle an exponential decaying process, recent data indicates that is not the case for longer observation time. The theoretical fractional differential calculus applied in the present work was able to describe this process at short and long time, explaining, in a single equation,  both exponential and non-exponential decay process. 
%{\bf Using a single equation  was  possible to model both of these behaviors  for the first time.}
%The fractional differential equation foi usada para modelar dados em duas regiões com comportamentos distintos.
%{\bf This paper will show} that fractional differential equation with noninteger order, such as
%{\bf $d^\alpha N(t)/dt^\alpha=-\lambda N(t)$ in which $0\geq \alpha\leq 1$, }can be used to interpolate
%luminescence decay process. %for all time.
%{\bf In order to} elucidate the meaning of fractional derivative, the Monte Carlo method was be used to simulate 
%a nonexponential decay process in which the probability of one particle undergoes emission is modified from 
%exponential decay process. 
%Metodologia%%%%%%%%%%%%%%%%%%%%%%%%%%%%%%%%%%%%%%%%%%%%%%%%%%%%%%%%%%%%%%%%%%%%%%%%%%%%%%%%%%%%%%%%%%%%%%%%%%%%
%\textcolor{red}{METODOLOGIA} 
The exact solution found by fractional model is given by an
infinite serie, the Mittag-Leffer function, 
with two adjusting parameters.  
%that describing presents asymptotic properties as observed in experimental data at long time.
To further illustrate this nonexponential behaviour and the fractional calculus framework, an stochastic analysis is also proposed.\\


\clearpage
\section{Introduction}
\noindent
%\textcolor{red}{CONTEXTUALIZAR}
%Unstable species decay
The question about validation of radioactive decay %exponential 
law has a long history and two reasons for this can be considered. One is that
%given: first
accurate experimental data at very long time are difficult to obtain. For example, for the $\alpha$-decay of  $^{\rm 8}$Be the deviation from exponential decay is expected after 30 half lives, when the signal intensity becomes of order $10^{-7}$ of the initial intensity \cite{KELKAR}. %{\bf
The other reason is that the exponential
decay law can be derived only as an approximate result of quantum mechanics theory. % \cite{PERES}. 
A rigorous quantum study
shows that at very short and very long
time unstable state decays with nonexponential dependence of time \cite{MERZBACHER,PERES}.%\\
%{\bf If nonexponential decay is confirmed for nuclear decay, this deviation will effect the radiocarbon dating, as suggested by Aston 
%\cite{ASTONa,ASTONb,NICOLAIDES}.}\\
%, and exponential law is .
%, and not a rigorous, result of
%quantum mechanics.
%}
%In 1980, Peres shown that decays starts as a quadratic function of time and ends as an inverse power law, in which,
%between these two extremes, the exponential day laws is approximately valid.
%{\bf A discussion about just what the quantum theory says, can be found in Peres work \cite{PERES}.}

%One should note that
%\noindent
%\textcolor{red}{FRONTEIRA DO CONHECIMENTO}
%Note that 
Until recently  the experimental searches for nonexponential behavior
%data to prove of
%has been
was
mainly focused on nuclear decay \cite{AVIGNONE,GODOVIKOV,SKOROBOGATOV},
in which, as discussed, the experimental measurements are difficult to perform accurately in such a long time.  If nonexponential decay is confirmed for nuclear decay, this deviation will affect 
%for example 
the radiocarbon dating, 
as suggested by Aston  \cite{ASTONa,ASTONb}.%,NICOLAIDES}. 
Experimental effort has now been directed to the study of luminescence decays of
%many dissolved
some organic
materials after pulsed laser excitation \cite{ROTHEa,ROTHEb}
and in this case the 
%In luminescence experiment the 
intensity data can be obtained with relatively less 
difficulty if compared with nuclear radiation decay.
%Using luminescence data,
%By way, the
%Rothe and coworkers[] presents a
The first experimental confirmation of the nonexponential decay law, using luminescence data,  was presented by Rothe and coworkers, in 2006 \cite{ROTHEb}.
These authors observed that, as expected,
initially luminescence decays obeys the radioactive decay law, but after some half lives %times
%the decay curve is slowly
the fluorescence decays significantly slower, turning into a different decaying law. %\\
In the study by Rothe and coworkers the experimental data were fitted to two different functions:
%Previous(DE QUEM) Experimental data are
%modelling into two regions
%collected into two regions 
%to facilitate the interpretation: % of the data:
%the first with exponential decay up to about 0.6ns, after which decay behavior changes to power law within the second region.
%the first was 
an exponential function up to about 0.6ns, and after 
%when
which it 
%thedecay behavior changes 
was used a power function.
% within nonexponential %the second 
%region.}%\footnote{Ele fez o ajuste em duas partes antes de 0.6ns usando $exp(-kt)$ e depois $1/t^n$}
%\textcolor{red}{LACUNA NA ÁREA}

This problem can be analysed in different way by using fractional calculus.
Recent studies %on this subject %in many fields of science,
have shown that fractional calculus is a good alternative mathematical tool
into many fields of science, such as the study of the growth of bacteria in culture media \cite{RIDA}, the mechanisms by which diseases spread \cite{POOSEH} and the kinetics of drug absorption \cite{DOKOUMETZIDIS}. % and nuclear decay \cite{CALIK}.
%In the study 
%  %performed 
%by Çalik, fractional order was found using as information just a half-life time. 
%nuclear physics \cite{CALIK}. %\footnote{Repetitivo com a introducao}
%However  
%Recently, the use the 
The use the fractional calculus was explored in nuclear decay by  
%\textcolor{red}{FALTA LIGAR! recententemente ...In the study 
%  %performed 
\c Calik\cite{CALIK}, in which a comparison between experimental data and theoretical model was made 
with %a few experimental 
%data 
%just 
just one half lives %time 
and always at short time when exponential behavior dominates.
%%%%%%%%%%%%%%%%%%%%%%%%%%%%%%%%%%%%%%%%%%%%%%%%%%%%%%%%%%%%
%questão: Mittag-leffler já foi usada antes
%\noindent
%\textcolor{red}{TERCEIRO RESULTADO DESTE TRABALHO: A equação de  Mittag-Leffer ajusta bem aos dados experimentais em tempos curtos e longos. Não existe na %literatura comparação semelhante mostrando que o ajuste é adequado para parte exponencial e não exponencial.}
%%%%%%%%%%%%%%%%%%%%%%%%%%%%%%%%%%%%%%%%%%%%%%%%%%%%%%%%%%%%%
%Therefore, the good fit between fractional model and experimental data at long time, 
%\textcolor{red}
%{when exponential decay law is not expected to work}, has not yet been %completely 
%established.%\\ %\footnote{Neste frase tento dizer que ML(ou fractional calculus) não foi usado ainda para explicar os dois comportamentos.}\\%\footnote{Achei vago aqui, que tal falar mais especificamente dos artigos que cita.} \\%In this paper fractional model will confronted with a wide range in time of experimental data.
%{\bf In 2012, Casasanta\cite{CASASANTA} presents modified Beer-Lambert law with derivative of fractional order, however no comparison with a experimental %data is carried out.
%Another work that use fractional differential equation into this context is performed by Çalik %work's
%\cite{CALIK}, however few data were used in your discussion.
%In the study 
%%performed 
%by Çalik, fractional order was found using as information just a half-life time. }
%{\bf dizer que esta equação já foi usada antes dentro deste contexto em...}
%\noindent
%\textcolor{red}{PROPOSTA}

In the present work 
%a discussion of the fractional calculus will be carried out.
%Then
a differential equation with fractional order will be presented as an attempt to generalize the model to
exponential and nonexponential
luminescence
decay of polyfluorene. %process. 
{For the first time, 
experimental data for
luminescence decays, %of polyfluorene, 
at short and long time, will be compared with the proposed model}
%\footnote{Nesta frase tendo dizer que isto não foi feito ainda para 
%dados de decaimento luminescente, na verdade para nenhum dado experimental com esta mudança de comportamento em tempos longos.}
%\textcolor{red}{\bf pelo primeira vez com o modelo proposto}\footnote{nesta frase tendo dizer que isto não foi feito ainda para 
%dados de decaimento luminescente, na verdade para nenhum dado experimental com esta mudança de comportamento},
%modelled with fractional differential equation. %and 
%{\bf Finally, the fractional model will confronted with a wide range in time of experimental data,  
%In this paper fractional model will confronted with a wide range in time of experimental data.
% proposed here.
%The main goal of this paper is to show 
indicating that fractional model is adequate to describe exponential and nonexponential chemical luminescence decay process. 
%, in wide time region,
Another issue to be considered here is how to define the probability in this luminescence decay process when described by
fractional differential equation. In a comparison with models of integer orders, 
%Monte Carlo simulation , will be considered %we will consider  
%the same algorithm used to simulate differential equation with integer order, in which
probability of $dN$ particles undergoing emission will 
here be
%now 
defined using the 
%infinite serie of 
Mittag-Leffer function. 
%\footnote{Tentei desvincular com o MC. Por que tem de falar com MC, a proposta \'e 
%estat\'{\i}sca.}
%\subsection{A incluir}
%{\bf Também não foi usado para decaimento luminescente antes.}\\

%\noindent
%{\bf Nenhum trabalho prático usa ML para explicar os dois comportamentos (só trabalhos matemáticos).}

%Finally, experimental data for
%luminescence decays of polyfluorene, at short and long time, will be modelled with equation proposed here.

%o caminho seguido aqui foi...

%as simulated a chemistry process describe by fractional differential
%equation.

%\section*{Theoretical background}
%\clearpage
\section{Fractional calculus background}

%A origem do cálculo fracionário remonta ao final do século XVII
%com a discussão entre Leibniz e L'Hospital sobre o siguinificado de  $d^{1/2}/dx^{1/2}$.
%Desta época até o momento atual, ilustres
%matemáticos como  Euler, Lagrange, Laplace, Fourier, Abel, dentre
%outros, tem contribuído para o desenvolvimento da teoria de cálculo fracionário. Para
%mais detalhes sobre o assunto veja o trabalho de dos Santos (dos SANTOS) e as referências
%contidas neste trabalho.

%In 1965, 
Fractional calculus has its roots in 
a discussion between Leibniz and L'Hospital about the meaning of $d^{1/2}f(x)/dx^{1/2}$ as described in a %historical 
book on the subject %\footnote{Conferir}  
\cite{1965}.
%From this time until now, the fractional calculus evolved in the research field of pure %mathematics and applied science.
%is a branch of mathematical pure anda applied that studies differential operators with non-integer orders.
Nowadays, fractional calculus is a branch of mathematics
analysis
that generalizes the derivative and integral
with noninteger order.%\\
%of a function to non-integer order.
%Recent studies on this subject, %in many fields of science,
%have shown that fractional calculus is a good alternative mathematical tool
%into many fields of science, such as the study of the growth of bacteria in culture media \cite{RIDA}, the mechanisms by which diseases spread %%%
%\cite{POOSEH} and the kinetics of drug absorption \cite{DOKOUMETZIDIS}.\footnote{Repetitivo com a introducao}

%To begin with  the definition of fractional
%integral which will be used to define the  fractional derivatives.
%For understanding
%to represent
%Therefore,  consider the integral operator defined by
%Para o entendimento do modelo proposto neste trabalho apresenta-se inicialmente
%uma motivação para a definição de uma
%derivada fracionária, partindo da definição de integral fracionária.
%Assim, considerando o operador integral $(Jf)(t)$ como
%\begin{equation}
%$(Jf)(t)=\int_0^t f(s) ds$.
%\label{int}
%\end{equation}
%\noindent
Theoretical description of fractional calculus may start by
integrating $(Jf)(t)=\int_0^t f(s) ds$ {with respect to $s$ by}
$m=(n-l)$ times from which one obtains \cite{PODLUBNYbook},
%have done the
%integration be $m$ times,
%the above equation %(\ref{int})
%can be written as follows \cite{PODLUBNYbook}
%by repeating $m$ times the operator
%tem-se então para a aplicação de $m$ vezes do operador pela esquerda a expressão

%\begin{equation}
%(J^mf)(t)
%%=(JJJJ...JJf)(t)
%=\frac{1}{\Gamma(m)}\int_0^t (t-s)^{m-1} f(s) ds
%\label{eqint}
%\end{equation}

\begin{equation}
(J^{(n-l)}f)(t)
%=(JJJJ...JJf)(t)
=\frac{1}{\Gamma(n-l)}\int_0^t (t-s)^{n-l-1} f(s) ds
\label{eqint}
\end{equation}

\noindent
with $\Gamma$ the gamma function.  On the other hand, the differential operator, $D$, can be further applied $n$ times to this integral to furnish,
%
%Equation (\ref{eqint}) is well defined for all positive $m$. %$m>0$
%Therefore, it is able to use Equation (\ref{eqint}) as definition of
%fractional integral, known as fractional integral of Riemann-Liouville $(J^\alpha f)(t)$, %in which
%$\alpha=m$ is the fractional order.
%%from which
%%A expressão acima é bem definida para $m$ não inteiro o que leva
%%a definição da integral fracionária de Riemann-Liouville $(J^\alpha f)(t)$ pela Equação %(\ref{eqint}), em que $m=\alpha$ é a %ordem fracionária (dos Santos).
%The fractional integral of Riemann-Liouville can be used as motivation to define the %fractional
%derivatives.
%
%If consider the fundamental theorem of calculus such as %in which
%$(DJf)(t)=f(t)$ then it can be possible to rewritten as
%$(D^mJ^mf)(t)=f(t)$ by finite induction method, where $m$ is a integer number %\cite{PODLUBNYbook}.
%In order to find a fractional
%derivatives definition, take $m=n-l$ and apply differential operator $D^l$ on both sides %of equation, thus
%it can be written as
%%$(D^nJ^{n-\alpha}f)(t)=(D^\alpha f)(t)$.
%% it can find .
%%A integral fracionária Riemann-Liouville será útil na definição de uma expressão para a %derivada fracionária
%%se levarmos em conta o teorema fundamental do cálculo, em que
%%\begin{equation}
%%(DJf)(t)=f(t)
%%\end{equation}
%%\noindent
%%Neste caso, por indução finita, tem-se (dos Santos)
%%\begin{equation}
%%(D^mJ^mf)(t)=f(t)
%%\end{equation}
%%\noindent
%%em que $m$ é um valor positivo e inteiro.
%%Considerando $m=n-\alpha$ e aplicando o operador $D^\alpha$ pela esquerda em ambos os %lados da equação, tem-se

\begin{equation}
(D^nJ^{n-l}f)(t)=(D^l f)(t)
\label{eqdif}
\end{equation}

\noindent
Therefore, for a non-integer number $l=\alpha$, 
{one obtains Riemann-Liouville fractional derivative. Another attempt to define was made by Caputo,  \cite{PODLUBNYbook}
%exchanging derivative and integral operator as
%strategic hypothesis,  one may write,}
as}
%
%Observe that on the left side of the Equation (\ref{eqdif}) is well defined even that l is %noninteger.
%Therefore, Equation (\ref{eqdif}) can be used as motivation to define the %Riemann-Liouville fractional derivative \cite{PODLUBNYbook}.
%%Note que o lado esquerdo equação acima é bem definida mesmo que l não seja um número %inteiro
%%A expressão acima é bem definida para $n$
%%inteiro maior que $\alpha$ e para qualquer valor de $\alpha$ positivo.
%If the order of derivation and integration is interchanged in Equation (\ref{eqdif}), %as
%%strategic,
%the result is
%Equation (\ref{eqdif}) can be written as
%,it can used the Equation (\ref{eqdif}) as motivation to defined the fractional derivative
%$(D^\alpha f)(t)$, as
%Portanto, pode-se usar
%a Equação (\ref{eqdif}) como motivação para definir a derivada fracionária $(D^\alpha f)(t)$, como

\begin{equation}
(D^\alpha f)(t)=(J^{n-\alpha}D^n)f(t)=\frac{1}{\Gamma(n-\alpha)}\int_0^t\frac{f^{(n)}(s)ds}{(t-s)^{\alpha-n+1}}
\label{caputo}
\end{equation}

\noindent
in which  $n\leq \alpha \leq n+1$, $\alpha \in \mathscr{R}$, $n \in \mathscr{N}$ 
and $f^{(n)}(s)=d^nf/ds^n$. 
In this case $\alpha$ is interpreted as the fractional derivative order.
%the fractional derivative becomes a well known Abel integral equation.
Equation (\ref{caputo}) is known as Caputo fractional derivative of order $\alpha$, that is in fact an Abel integral equation \cite{abel}. Dealing with fractional derivative and inverting Abel integral equation, an ill-posed problem, are equivalent problems. %\\
%\noindent
%in which the order of derivation and integration was changed as
%strategic, %therefore Caputo fractional derivative
%becoming a well known Abel integral equation. The Caputo and Riemanm definitions are equivalent under some conditions \cite{PODLUBNYbook}.
%as strategy of calculus.
%Therefore, Caputo fractional derivative is defined by Abel integral.
%form
%conhecida como derivada fracionária de
%ordem $\alpha$ segundo Caputo (dos Santos), em que a ordem entre derivada e integral foi alterada em relação a Equação
%\ref{eqdif}). Na Equação (\ref{caputo}) tem-se que $f^{(n)}(s)=d^nf/ds^n$.

The Caputo fractional derivative
is a nonlocal operator for it  depends on the strain history from $0$ to $t$.
This should be contrasted with a derivative of integer order, a clear local operator.
%are non-local operators because they are defined using integrals.
Consequently the fractional derivative in time contains information about the function at earlier points, an effect known as
memory effect which will %be shown
explored along the current work. This definition to fractional derivative  can be used to model the rate change because 
%Caputo fractional 
derivative of a constant is zero, unlike Riemann-Liouville definition. Therefore, henceforth make the assumption that $D^\alpha f(t)$ is Caputo 
fractional derivative.
%, which means that can be used to model
%in this paper with Monte Carlo approach.

%\noindent
% In contrast to differential operators of integer order
%Esta é a principal vantagem das derivadas de ordem fracionária em comparação com
%as derivadas de ordem inteira, onde tais efeitos são negligenciados. O operador diferencial com ordem fracionária é um operador não-local em quanto o operador %de ordem inteira é um operador local. A propriedade da não localidade de um operador
%consistem do fato de que o próximo estado de um sistema não só depende de seu estado atual, mas também de todos
%os estados anteriores a partir do estado inicial.

%\subsection*{Laplace transform method in fractional differential equation}
%\subsubsection*{A generalized decay law}
%\clearpage
\section{A generalized decay law}

An unimolecular  
process is described by %following differential equation
%Como discutido na introdução, em geral, o decaimento radioativo é bem descrito por uma equação diferencial de ordem inteira (NOVKOVI\'C; NORMAN)

\begin{equation}
dN(t)/dt=-k N(t),
\label{edo}
\end{equation}

\noindent
with the solution given by
%cuja solução é dada por
$N(t)=N(0)e^{-k t}$. The constant $k$ is the decay constant,
$N(t)$ the number of species
%nuclei
present at a time $t$ and $N(0)$ is the number of particles at initial time.  This is an example of a process described by an integer order derivative.  %In the problem to be considered here, fluorescence process, the quantity $N(t)/N(0)$ represents the relative
%emission intensity at time t. 
Although the unimolecular model is the one normally used to describe the emission intensity $I(t)=N(t)/N(0)$ of a luminescence process, recent 
%\footnote{Certo?} 
experimental data evidenciate that can not be the case for long observation time. %\\

{As luminescence decays behavior changes with time, it we will assume that the memory effect have an important role in decays process. 
One way to include this effect  is  to use fractional derivative order in the model.}
%effect is to  }
Investigation of the
%Modelling 
chemical luminescence process by a fractional derivative order has to start by the Caputo derivative,
%is also a commom procedure. 
%As previously discussed we 
%propose to consider a generalized differential equation with noninteger order, such as
%can be able to interpolate
%Na tentativa de descrever o processo
%para pequenos e longos tempos de observação, o presente estudo sugere uma
%generalização da Equação (\ref{edo}) para uma ordem não inteira, tal como

\begin{equation}
D^\alpha I(t) %=-k^\alpha I(t)
=-\lambda I(t)%\footnote{Temos que destacar a dimens\~ao do $\lambda$}
\label{edf}
\end{equation}

%\newline
\noindent
and, as to be discussed, will correct to describe {luminescence experimental data in both regions:
exponential and nonexponential part. This is the first study in the literature 
of such a data using fractional analysis.} %\footnote{Sim, este \'e o primeiro trabalho que usa c\'alculo fracion\'ario para descrever dados experimentais 
%em que se observa esta mudança de comportamento.}
%From Equation (\ref{edf}), it follows that $\lambda=k^\alpha$ {\bf and  $N(0)=1$. If  }
%and $0\leq \alpha \leq 1$. If 
If fraction order is equal a one 
%{\bf and  $N(0)=1$} 
recovers the usual description of this process, equation (\ref{edo}). %, {\bf with $I(t)=N(t)$}.
Henceforth we will assume that $N(0)=1$ and $N(t)=I(t)$.

%%%%%%%%%%%%%%%%%%%%%%%%%
%\clearpage
\subsection*{Laplace transformation and fractional derivative}
%\footnote{Mudar notação do $l$ e $1$ que está confuso}
%Laplace transform is a method to handle integration of fractional differential equation, in the same way that is
%used in differential equation with integer order. Therefore, instead of solving directly %Equation (\ref{edf})
%for $N(t)$, it can be written a
%new equation for $n(s)$, in which  $n(s)=\mathfrak{L}[N(t)]$ and $\mathfrak{L}$ is Laplace %operator.
%Once found $n(s)$, then $N(t)$ can be determined whenever $\mathfrak{L}^{-1}[n(s)]$ is %known.
%Once we find Y(s), we inverse transform
%to determine y(t).
%A transformada de Laplace representa uma estratégia %são uma ferramenta
%importante para a determinação das soluções de equações diferencias fracionárias assim como para o caso de equações diferencias de ordem inteira.

{{Consider the initial value problem %for a homogeneous fractional differential equation 
(\ref{edf}), 
with $n-1<\alpha\leq n$
%with $t>0$
%under 
%non-zero
and 
initial conditions given by \mbox{$[D^{l} N(t)]_{t=0}=b_l$} where $l=1,2,...,n$.} 
Applying Laplace transform to both sides of the
differential \hspace{.25cm} equation\hspace{.25cm}  (\ref{edf}),\hspace{.25cm}  that\hspace{.25cm}  is\hspace{.25cm} 
\mbox{$\mathfrak{L}[(D^\alpha N)(t)]=\mathfrak{L}[-\lambda N(t)]$}}, 
%O método consiste em aplicar o operador $\mathfrak{L}$ em ambos os lados da Equação
%({edf}), então
%\begin{equation}
%\mathfrak{L}[(D^\alpha N)(t)]=-\lambda \mathfrak{L}[N(t)]
%\end{equation}
%\noindent
using the property\cite{PODLUBNYbook,podart}%\footnote{Procurar referência}
%\footnote{uso a eq 4.1 da pag 138 do livro. Acho que esta alteração responde o comentário inicial do revisor 1 sobre a organização do artigo (comentários %geral do revisor 1).}  
%$\mathfrak{L}[(D^\alpha N)(t)]=s^\alpha\mathfrak{L}[N(t)]-s^{\alpha-1}N(0)$,
%$%\begin{equation}
%\mathfrak{L}[(D^\alpha N)(t)]=s^\alpha\mathfrak{L}[N(t)]-s^{\alpha-1}N(0)
%$%\end{equation}

\begin{equation}
%\mathfrak{L}[(D^\alpha N)(t)]=s^\alpha\mathfrak{L}[N(t)]-
%\sum_{k=0}^{n-1}s^k [D^{\alpha-k-1} N(t)]_{t=0}
%-s^{\alpha-1}N(0)
\mathfrak{L}[(D^\alpha N)(t)]=s^\alpha\mathfrak{L}[N(t)]-
\sum_{k=0}^{n-1}s^{\alpha-k-1} [D^{k} N(t)]_{t=0}
\end{equation}

\noindent
%with $n-1<\alpha\leq n$  
%\footnote{Neste caso, $n-1<\alpha\leq n$. Falta incluir isto no texto.}
%\hspace{.1cm} 
and after some simple algebraic manipulation, one obtains
%can be  developing one obtains
%um resultado conhecido da literatura (dos Santos), tem-se

\begin{equation}
%\mathfrak{L}[N(t)]=\frac{s^{\alpha-1}}{s^\alpha+\lambda}
\mathfrak{L}[N(t)]=\sum_{l=1}^{n}\frac{b_ls^{\alpha-l}}{s^\alpha+\lambda}
\label{laplace}
\end{equation}

\noindent
where $l=k+1$ and $b_l=[D^{l} N(t)]_{t=0}$.
%
%\noindent
%The next step consist to find the
The function that satisfies {each term} of Equation (\ref{laplace}) is the Mittag-Leffer function  %\footnote{Joao, acha que devo detalhar mais isto?}
with two parameter, $E_{\alpha,\beta}(-\lambda t^\alpha)$, {{such which} \cite{PODLUBNYbook,podart} }
\begin{equation}
%\mathfrak{L}[E_{\alpha,1}(x)]=\frac{s^{\alpha-1}}{s^\alpha+\lambda}
\mathfrak{L}[t^{\beta-1}E_{\alpha,\beta}(-\lambda t^\alpha)]=\frac{s^{\alpha-\beta}}{s^\alpha+\lambda},
\end{equation}
%\bf colocar a condi\c c\~ao $ Re(s)>|\lambda|^{1/2}$. }
%\footnote{artigo do Podlubn, eq. 1.15 pag.7}
%with one parameter, $E_{\alpha,1}(x)$, 
%in which $x=-\lambda t^\alpha$ \cite{PODLUBNYbook}. 
% \\
%\noindent
Therefore, solution of
Equation (\ref{edf}) is given by %\footnote{Usei a propriedade 32.3 do Murray, 'propriedade da soma das inversas'}
%Neste caso, a solução é dada pela função cuja transformada de Laplace é
%$s^{\alpha-1}/(s^\alpha-\lambda)$. Considerando que
%\begin{equation}
%\mathfrak{L}[E_{\alpha}(-\lambda t^\alpha)]=\frac{s^{\alpha-1}}{s^\alpha+\lambda}
%\end{equation}
%\noindent
%outro resultado conhecido da literatura, tem-se, finalmente,

\begin{equation}
%N(t)=\sum_{l=1}^n b_lt^{\beta-l}E_{\alpha,\beta}(-\lambda t^\alpha)
N(t)=\sum_{l=1}^n b_lt^{l-1}E_{\alpha,l}(-\lambda t^\alpha)
\label{final}
\end{equation}

\noindent
%{\bf in which $\beta=\alpha-l+1$\footnote{Pensar se vale a pena colocar!}. 
If consider $\alpha$ between 0 and 1, %$0<\alpha<1$ 
and initial condition $b_1=1$, %and $b_l=0$ with l=2,3,...,n, 
one obtains

\begin{equation}
%N(t)=b_1t^{\beta-1}E_{\alpha,\beta}(-\lambda t^\alpha)
%N(t)=t^{\beta-1}E_{\alpha,\beta}(-\lambda t^\alpha)
N(t)=E_{\alpha,\beta}(-\lambda t^\alpha)
\label{final}
\end{equation}

%\noindent
%$x=-\lambda t^\alpha$.
%and 
\noindent
%{\bf with $\beta=1$}. 
{Equation (\ref{final}) represents} a generalized decay law for the present study and which  
%{Portanto, chego na equação que estava usando!}
%This generalized law
%The equation \ref{final}
%that 
will be used to describe
experimental data of luminescence intensity emission of a organic molecule, after pulsed laser irradiation. 
{The quantities, $\alpha$ and $\beta$ are considered as parameters that have to be estimated. Equation  (\ref{final}) 
was previously used to study of pharmokinetics\cite{DOKOUMETZIDIS} and nuclear decay\cite{CALIK} with $\beta=1$. }
%$E_{\alpha}(x)$ is the Mittag-Leffer function.
The proposed model reproduces purely exponential behavior when $\alpha=\beta=1$. 
%{\bf with initial condition $N(0)=1$}.
%, since in this case
%$E_{1,1}(x)=e^x$.
%\subsection*{Mittag-Leffer function}
%------------------

\subsection*{Mittag-Leffer function with two parameter}

The Mittag-Leffer function {with two parameter} can be seen as a generalization of the exponential function
%is 
defined by the following infinite power series \cite{Mainardi}
%A função de Mittag-Leffer
%, definida pela série infinita

\begin{equation}
E_{\alpha,\beta}(x)=\sum_{k=0}^\infty \frac{x^k}{\Gamma(\alpha k+\beta)}
\label{ml2}
\end{equation}

\noindent
{with $\alpha,\beta > 0$,} in which $E_{\alpha,\beta}(x)=e^{x}$ when $\alpha=\beta=1$.  
%\sout{In 1953, by Agarwal, 
%a new generalization of the Mittag-Leffer is obtained by replacing the constant 1 in the 
%argument of the Gamma function by an arbitrary parameter $\beta$, as follows \cite{Mainardi}} 
%\begin{equation}
%\sout{E_{\alpha,\beta}(x)=\sum_{k=0}^\infty \frac{x^k}{\Gamma(\alpha k+\beta)}}
%\label{ml}
%\end{equation}
%\footnote{Caiu muito depressa na funcao ML. De mais informacoes teoricas. Tente aproveitar o texto logo abaixo desse footnote.}\\
%\noindent
%\sout{ in which $E_{\alpha,1}=E_\alpha$.}%\\
%A description of the most important properties of the Mittag-Leffler function, with one or two parameter, can be found in reference \cite{Mainardi}.
%For our work the most interesting properties are asymptotic behavior at 
%The Mittag-Leffer function with two parameter is defined by the following infinite power series \cite{MITTAG}
%A função de Mittag-Leffer
%, definida pela série infinita
%\begin{equation}
%E_{\alpha,\beta}(x)=\sum_{k=0}^\infty \frac{x^k}{\Gamma(\alpha k+\beta)}
%\label{ml}
%\end{equation}
%\noindent
%The Mittag-Leffer functions can be seen as a generalization of the exponential function
%For $\beta=1$, 
%\noindent
This function  have interesting asymptotic properties, which was
studied by Mainardi \cite{Mainardi}.
%It is shown
In his work he showed
that for $t\rightarrow 0$ Mittag-Leffer function {with two parameters} converges to
exponential function whereas to $t\rightarrow \infty$ converges to $1/t^{\alpha r}$, therefore
%These asymptotic behavior are rewritten here for clarity
%The asymptotic behavior of Mittag-Leffer function
%tem um papel central no cálculo de ordem fracionária e propriedades assintóticas interessantes para
%descrição do fenômeno de decaimento radioativo. (MAINARDI)

\begin{equation}
E_{\alpha,\beta}(-t^\alpha)\approx \left\{
\begin{array}{ll}
\exp\left[-\frac{t^\alpha}{\Gamma(\beta+\alpha)}\right], \mbox{$t\rightarrow 0$}\\
\sum_{r=1}^\infty(-1)^{r-1}\frac{t^{-\alpha r}}{\Gamma(\beta-\alpha r)}, \mbox{$t\rightarrow \infty$}
%\approx
%\frac{t^{-\alpha}}{\Gamma(1-\alpha)}
\end{array}\right.
\label{serie}
\end{equation}

\noindent
%in which $x=-t^\alpha$. 
As these asymptotic behavior is similar to that observed in the Rothe experimental data, 
 motivated us to study the fractional calculus in luminescence decay of polyfluorene. %A description of the some %most important 
 %properties of the Mittag-Leffler function, with one or two parameter, can be found in reference \cite{Mainardi}.

%Therefore the Mittag-Leffer function can be used to interpolate experimental data at exponential and
%nonexponential region.
%The proposed model reproduces purely exponential behavior when $\alpha=\beta=1$, in this case
%$E_{1,1}(x)=e^x$.
%O comportamento assintótico da função de Mittag-Leffer %, conforme discutido na seção de Metodologia,
%é similar ao
%ao observado por Kelkar
%(KELKAR) para a descrever o decaimento-$\alpha$ do $^8Be(2^+)$.
%Portanto, a função de Mittag-Leffer
%é adequada para interpolar o decaimento exponencial em tempos muito pequenos, % tempos,
%enquanto que também é capaz de interpolar um decaimento mais suave
%para tempos longos de observação.
%O modelo proposto também se ajusta aos casos de decaimento puramente exponencial, quando $\alpha$=1 tem-se %$E_{\alpha,1}(x)=e^x$ e o resultado previsto pela Equação (\ref{edo}).%\\

\subsection*{Monte Carlo Method}
The Monte Carlo method \cite{num} is employed to simulate the luminescence decays of organic substance after pulsed laser excitation. The probability of emission,
$P=dN/N$, can be obtained by differentiating Equation (\ref{final}) with respect to $t$,  
%cccccNao teria de apresentar o dN/N ?ccccccc
%we obtain %one obtains

\begin{equation}
dN=-\lambda \alpha t^{\alpha-1} dt \frac{dE_{\alpha,\beta}(x)}{dx}
%\frac{dN}{dt}=\frac{dE_{\alpha,\beta}(x)}{dx}\frac{
\label{monte}
\end{equation}

\noindent
in which \cite{PODLUBNYbook,podart}

\begin{equation}
\frac{dE_{\alpha,\beta}(x)}{dx}=\frac{E_{\alpha,\beta-1}(x)-(\beta-1)E_{\alpha,\beta}(x)}{\alpha x}
\end{equation}

\noindent
and $x=-\lambda t^\alpha$. 
Under this considerations, the probability $P$ can be found, such as

%\begin{equation}
%\frac{dN}{N}=-
%\lambda  dt \left[t^{\alpha-1}  
%\frac{E_{\alpha,\beta-1}(x)-(\beta-1)E_{\alpha,\beta}(x)}{ x E_{\alpha,1}(x)}\right]
%\label{prob}
%\end{equation}

%\begin{equation}
%\frac{dN}{N}=
%\lambda  dt \left[t^{-1}  
%\frac{E_{\alpha,\beta-1}(x)-(\beta-1)E_{\alpha,\beta}(x)}{ \lambda E_{\alpha,\beta}(x)}\right]
%\label{prob}
%\end{equation}

\begin{equation}
\frac{dN}{N}=
t^{-1} dt \left[  
\frac{E_{\alpha,\beta-1}(-\lambda t^\alpha)}{ E_{\alpha,\beta}(-\lambda t^\alpha)}-(\beta-1)\right]
\label{prob}
\end{equation}

%
%\begin{equation}
%dN=\lambda \alpha t^{\alpha-1} dt \frac{dE_{\alpha,\beta}(x)}{dx}
%\label{monte}
%\end{equation}
%
\noindent
and simulation can be carried out by Monte Carlo approach.
The proposed equation reproduces
the exponential
behavior when $\alpha=1$ and $\beta=1$, for in this case $P=\lambda dt$ 
whereas \mbox{$E_{1,0}(-\lambda t)=-\lambda t E_{1,1}(-\lambda t)$}. 
The result obtained by Monte Carlo method will be compared with
those obtained from Equation (\ref{final}).

\section{Results and discussions}

\subsection*{Fractional calculus modelling}

%\noindent
%\textcolor{red}{REVISITANDO O ARTIGO (PONTOS CHAVES)}

\begin{center}
\begin{figure}[h]
\centering
\includegraphics[scale=1]{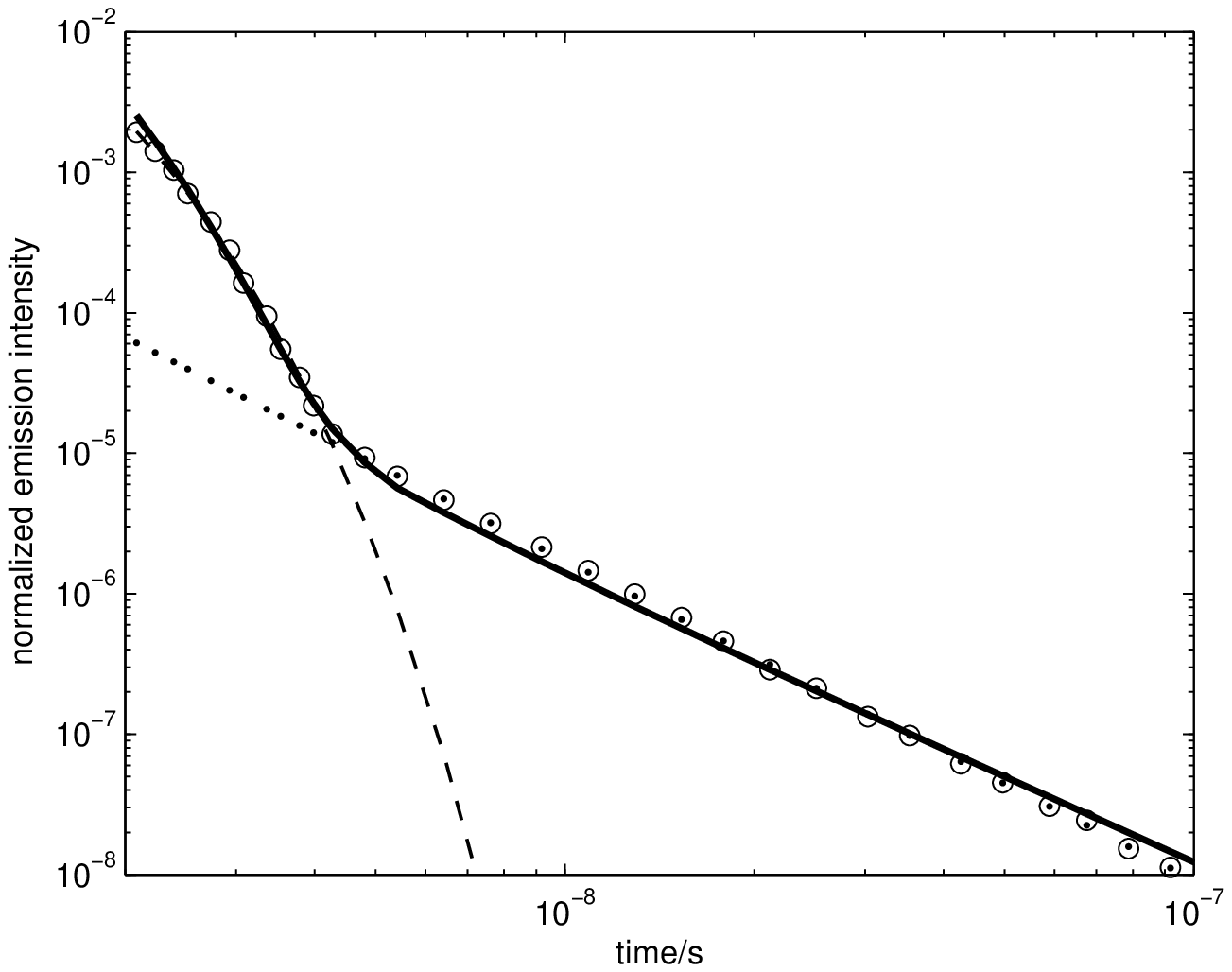}
%%C:\Users\nelson\Documents\My Dropbox\WORK\PQUISA\cinfra\main
%%ajustaprl.m
\caption{Experimental data for luminescence decay as open circle symbols, taken
 from the Rothe's work \cite{ROTHEb}. Exponential and power law are indicated by
dashed and dotted line, respectively. The continuous line represents the result obtained by 
the present work.}
%fractional differential equation. }
\label{figure1}
\end{figure}
\end{center}

Rothe and coworkers studied luminescence decays of polyfluorene after pulsed laser excitation
for a time enough to observe nonexponential behavior \cite{ROTHEb}. Figure \ref{figure1} shows their experimental results together
with two adjustments to the experimental data: one of them for the exponential decay region and the other for nonexponential one.
Nonexponential behaviour presents a decay law as a power law, in which emission intensity at time $t$ is proportional to $t^{-n}$. {For polyfluorene 
Rothe and coworkers determined $n=2.1$, 
by a linear adjustment of ($\ln N(t)\times \ln(t)$), at long time.  }
%The values of $n$ for other substances was found to be between $2.1$ and $4.1$.}%\\
%TIREI
%
%$n$ 
%value found by Rothe was 2.1. %is closely  
%exponent found by Rothe was $n$=2.1.
% with observed exponents 
%between 2.1 and 4.1 for other substances.cccc confuso, exponents...exponents ...cccccc

%\noindent
%\textcolor{red}{PRIMEIRO RESULTADO DESTE TRABALHO: O $\alpha$ deve ser muito próximo de 1! Conclusão deste trabalho muito diferente do Çalik, ele usa %alpha 0,5, 0,3 etc! E ele está errado em usar $\alpha$ longe de 1. } \footnote{Aqui ficou claro para mim: o Calik fez a derivada fracionaria entao, mas %nao usou ML. Devemos pensar em mudar o titulo do artigo para dar essa informacao. A two parameter Mittag-Lefer function to describe nonexponential %chemical effects }

The time in which a change in the decay behavior occurs
% have to turning of behavior
can be defined by intersection of exponential law and power law, as shown {in Figure \ref{figure1}}.
For polyfluorene substance this times was determined to be about 2.7 ns from the initial time, or about 11 half lives after ($t_{1/2}$=0.25 ns). 
%Therefore, the nonexponential behavior is observed at longer time, as a result the fractional order closed to integer order so as to guarantee a exponential behavior until many life-times.  
In the study 
%performed 
by \c Calik\cite{CALIK} {on nuclear decay}, the fractional order was found using 
only radiation intensity at one half life, 
%data 
%for just one {\bf time point, the half-life time}, 
and in this case it was found fractional order %much 
smaller than 1. 
Previous result can not explain the behavior of nuclear decay curve off half life.  
Here, the whole decay curve, with experimental data at short and long time, was %modelled 
fitted by fractional calculus  and not just one single experimental data such as radiation intensity at half life. %\\

{The code developed by Podlubny \cite{podmat} was used in this paper to determine the Mittag-Leffler function. }
%\sout{This routine validate in previous work \cite{artigonovo} provide Mittag-Leffler function with accuracy desired}.
Initially the behaviour of luminescence decay is exponential, therefore it is expected that 
$\alpha$ value is close  to 1. 
%\textcolor{red}{However should be different from 1 \sout{in order that} for the model can predict non-exponential behavior 
%at long time.} \sout{As we can see} 
As observed by the asymptotic series (\ref{serie}), the choice of $\alpha$ and $\beta$ parameters  
determines the relative weight of each of the infinite series. 
%Para garantir que a fun\c c\~ao de Mittag-Leffler 
%preserve o comportamento exponencial
%por um algum tempo, \'e necess\'ario que alpha seja muito pr\'oximo de 1 mas nessariamente diferente de 1 para 
%prever o comportamento n\~ao-exponencial em tempos longos de observa\c c\~ao.
% O par\^ametro $\beta$ permite controlar o peso dos 
%diferentementes termos da s\'erie 
%assint\'otica }
%%{\bf falar sobre o método para o cálculo da função de ml}\cite{podmat}
%{\bf prever o Beta usando a série}
%
%\begin{equation}
%E_{\alpha,\beta}(-t^\alpha)\approx 
%\frac{t^{-\alpha}}{\Gamma(\beta-\alpha)}-\frac{t^{-2\alpha}}{\Gamma(\beta-2\alpha)}+\frac{t^{-3\alpha}}{\Gamma(\beta-3\alpha)}-...
%\sum_{r=1}^\infty(-1)^{r-1}\frac{t^{-\alpha r}}{\Gamma(\beta-\alpha r)}, 
%\label{serie}
%\end{equation}
%where \mbox{$t\rightarrow \infty$}.
%
%\noindent
For example, if $\alpha=\beta$ the first term of the series is 
%\sout{null} 
zero, thus tbe second term  takes on a more important role, 
whereas the other terms tend to zero faster. Therefore  $E_{\alpha,\alpha}(-t^\alpha)\propto t^{-2\alpha}$ when $t\rightarrow \infty$, 
and finally $n\approx 2\alpha$. By previous supposition $\alpha\approx 1$ and ,therefore, $n\approx 2$ 
%\sout{which is case of} 
which is the case for
polyfluorene.
%{\bf Por exemplo, fazendo $\alpha=\beta$, temos que o primeiro termo da s\'erie \'e zero, e o segundo termo passa a ter uma 
%contribui\c c\~ao mais significativa para a s\'erie, considerando que os outros termos tendem a zero repidamente, 
%temos  $E_{\alpha,\alpha}(-t^\alpha)\propto t^{-2\alpha}$. 
%Portanto, 
%achamos $n\approx 2\alpha$. Como n=2.1 para a mol\'ecula de Poplyxxx tem-se que $\beta \approx 1$ tendo em conta que $\alpha\approx 1$.
{These values of $\alpha$ and $\beta$ were used as initial choice}.
% Estes valores 
%foram usados como chutes iniciais na otimiza\c c\~ao da fun\c c\~ao.} 
%= 1.998$ when $\alpha=0.999$. Se 

%\begin{equation}
%E_{\alpha,\beta}(-t^\alpha)\alpha
%t^{-2\alpha}
%-\frac{t^{-2\alpha}}{\Gamma(\beta-2\alpha)}
%\sum_{r=1}^\infty(-1)^{r-1}\frac{t^{-\alpha r}}{\Gamma(\beta-\alpha r)}, 
%\end{equation}

{In figure \ref{figure1}} continuum line represents the interpolation by
%fractional differential
equation (\ref{final}),
which was obtained with a value of $\alpha=\beta=0.999$ and $\lambda=2.8\times 10^{9}$ ns$^{-1}$.
%The code developed by Podlubny provide Mittag-Leffler function with accuracy desired
The %estimation of the
model parameters, $\alpha$, $\beta$ and $\lambda$,
were determined using Simplex method \cite{num} with the objective function defined by sum of squared
difference between experimental and calculated values.
As can be seen,
the Mittag-Leffer function with two parameters reproduce
experimental data with good agreement at short and long time.
%The theoretical results in Figure \ref{} was obtained with a value of $\alpha=\beta=0.999$ and $\lambda=$.
%Therefore,
{In other words,
with Equation (\ref{final}) it was possible to reproduce, using only one function,
%into one function,
the two different behavior: exponential and nonexponential.}
% the quality
The goodness of fit can be evaluated by the 
relative error between experimental data and
predict results, which is less than 1.5\%. {Calculated data by fractional model 
were adjusted by two linear equation, at short ($\ln N(t) \times t$) and long  ($\ln N(t) \times \ln t$) time.
The parameter $\tau=1/\lambda=0.36$ ns and $n=2.1$ %, which was found through Equation (\ref{final}),
also are in excellent agreement with those obtained using experimental data: %by Rothe and coworkers: 
$\tau=0.35$ ns and $n=2.3$. } {In figure \ref{figure1}, the dashed line  represents the result obtained from 
equation (\ref{edo})}.
%\footnote{Aqui fiquei confuso. Ate nesse ponto achava que ML com dois parametros era a ideia original do artigo. Parei aqui, MC vou deixar para quando %limparmos mais o artigo.}

%{\bf In 2012, Casasanta\cite{CASASANTA} presents modified Beer-Lambert law with derivative of fractional order, however no comparison with a experimental %data is carried out.
%Another work that use fractional differential equation into this context is performed by Çalik }%work's
%%\cite{CALIK}, however few data were used in your discussion.
%%In the study 
%%%performed 
%%by Çalik, fractional order was found using as information just a half-life time. }

%%%%%%%%%%%%%%%%%%%%%%%%%%%%%%%%%%%%%%%%%%%%%%%%%%%%%%%%%%%%%%%%%%%%%%%%%%%%%%%%%%%%%%%%%%%%%%%%%%%%%%%%%%%%%%%%%%%%%%%%%%%%%%%%%%%%%
\subsection*{Monte Carlo simulation}
%
%{\bf ccccc tireiAt this point, an important issue involves the question of what is the %meaning of fractional derivatives.
%One possible answer can be given by Monte Carlo simulation.
%In this paper, the Monte Carlo method was used to simulate luminescence decays when %described by
%differential equation with fractional order. }\\
%
The Monte Carlo method was further used to simulate and elucidate luminescence decays when described by
differential equation with fractional order.
In this simulation to radioactive decay
the probability that a element undergoes
changes in time $t$ is given by Equation (\ref{prob}).
%,  {\bf which is presented here for the first time.} 

%\noindent
%\textcolor{red}{QUARTO RESULTADO DESTE TRABALHO: O cálculo da probabilidade da forma proposta funciona na simulação de Monte Carlo!}

\begin{center}
\begin{figure}[h]
\centering
\includegraphics[scale=1]{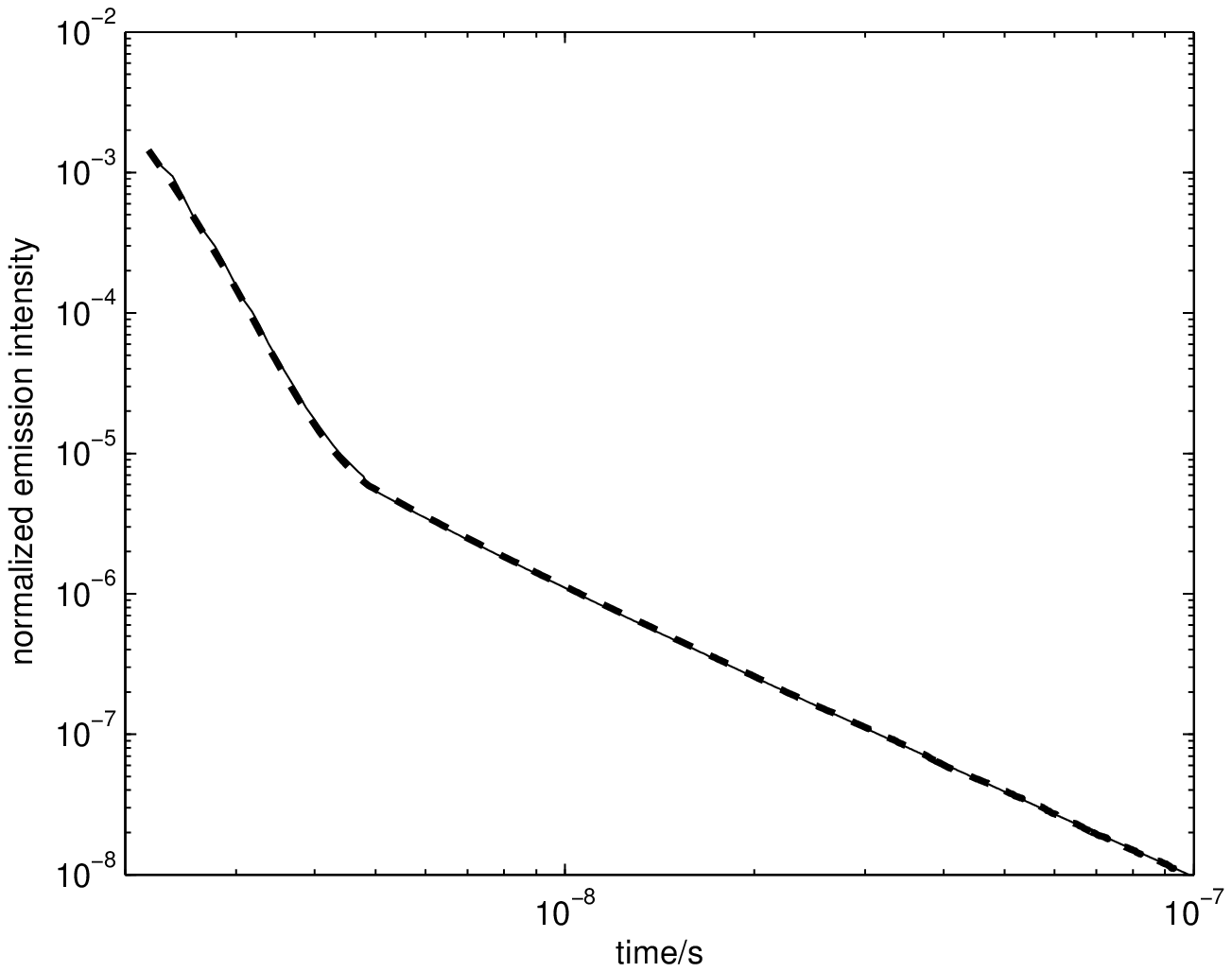}
%%C:\Users\nelson\Documents\My Dropbox\WORK\PQUISA\cinfra\main
%%
\caption{Present theoretical result (continuous line), together with
result obtained by Monte Carlo method (dashed line).}
\label{figure2}
\end{figure}
\end{center}

The simulation was performed with N=50 000 000 and Mittag-Leffer function was calculated by Podlubny
routine\cite{podmat} with accuracy of $10^{-12}$. {Figure \ref{figure2}} shows the intensity decay as function of time
obtained by Monte Carlo
method together with the results obtained by equation \ref{final}.
The relative error observed was less than 0.7 \%.
%experimental results.
{Therefore, one may conclude that the probability defined by Equation (\ref{monte}) is
adequate to describe luminescence decays at wide range of time.} %at long time.

%%%%%%%%%%%%%%%%%%%%%%%%%%%%%%%%%%%%%%%%%%%%%%%%%%%%%%%%%%%%FALTA
%detalhes do monte carlo

%\noindent
%\textcolor{red}{QUINTO RESULTADO DESTE TRABALHO - Interpretação da equação fracionária pela simulação proposta: a probabilidade não é mais constante!}

%The probability as function of time is shown {in Figure \ref{figure3}}. 
For differential equation
with integer order the probability as function of time is constant during all
%luminescence decay
process. On the other hand, when noninteger order
is used the probability changes with time. Therefore, when considering derivatives with noninteger order, the probability that $dN$ particles undergos emission not only depend of number of particle at time $t$, but on the whole process.
%%%%%%%%%%%%%%%%%%%%%%%%%%%%%%%%%%%%%%%%%%%%%%%%%%%%%%%%%
%%%%%%%%%%%%%%%%%%%%%%%%%%%%%%%%%%%%%%%%%%%%%%%%%%%%%%%%%
The reason for this is that fractional derivative is defined by integral equation as shown previously.
%{\bf Este resultado está de acordo como o que foi apresentado sobre....}
%This is an important effect taking into account with fractional order.
%The memory effect is
%taking into
%account memory mechanism in luminescence decays, it is
This effect is well known in other problems with fractional derivative and named %labelled
as memory effect \cite{DOKOUMETZIDIS,POOSEH,RIDA}.
\section{Conclusions}

%\noindent
%\textcolor{red}{Começar específico e ir para o geral!}

%The luminescence decay is often assumed as exponential, but was
%  %it is
%clearly shown
%by Rothe \cite{ROTHEb}
%  %in xxxxx study
%that there existis a time regime where decay process exhibits
%a nonexponential time dependence.
%  %that perfectly coincides with the straight line at long enough time.
%  % with slope $-\lambda$.
%For polyfluorene, the decay process at long times is characterized by
%a  $t^{-n}$ dependence, in which $n\in \mathscr{R}>0$.

%Often two equation are used to modelling these two different regions: exponential and nonexponential.
%  %The main purpose of this paper is to find the
%However, %this paper
%our model describes
%the decay of luminescence using just one equation.
%The %fractional
%differential equation with fractional order was used to
%modelling experimental emission in polyfluorene after pulsed laser excitation.

%\noindent
%\textcolor{red}{APRESENTAR OS PRINCIPAIS RESULTADOS}

%Our model with fractional order evaluated experimental data with relative error less than 1.5\%. %, in which
%$\tau=1/\lambda=0.36$ ns and $n=2.1$. %These results are in excellent agreement with those obtained by Rothe and coworkers.
%From Mittag-Leffer function, found by fractional model, was able to interpolate experimental data at all time.
The solution found by fractional model was given by Mittag-Leffer function with two parameters, which was able to interpolate experimental data 
between 1 ns and 100 ns, unlike previous studies in which few data at short time were used. 
For polyfluorene molecule 
this interval is so long that it was possible to observe a change in the exponential behavior.  
%from exponential behavior for a power law
Our model with fractional order coincide with experimental data within a relative error less than 1.5\%. 
%at all time.
The parameters used in Mittag-Leffer function fit were $\alpha=\beta=$0.999 and $\lambda$=0.36.

%The estimation of the model parameters, $\alpha$ and $\lambda$,
%were determined using Simplex method \cite{simplex} with the objective function defined by sum of squared
%difference between experimental and calculated values.
%was performed by minimization of an objective function using Simplex method\cite{simplex}.
%As can be seen,
%the Mittag-Leffer function with two parameters reproduce

The probability of $dN$ particles undergoing emission was here defined for the first time using Mittag-Leffer function with two parameters, and the 
Monte Carlo method was employed to simulate the luminescence decays of polyfluorene molecule. 
A good agreement between exact result and Monte Carlo approach was also obtained, with a 
%The relative error found
relative error %found so is 
less than 0.7\%.
%with relative error less than x\%.
This result shows that our definition of
probability %found
%of $dN$ particles undergoes emission 
is adequate to describe experimental data  in a wide range of time, 
taking into account memory effect, by using noninteger order in differential equation. 

\section*{Acknowledgments}

We would like to thank CNPq and FAPEMIG for their financial support.

\clearpage

\clearpage


\clearpage

\begin{thebibliography}{}

\bibitem{KELKAR} N. G. Kelkar, M. Nowakowski, K. P. Khemchandani, 
%Hidden evidence of nonexponencial nuclear decay.
%{\it Phys. Rev. C}, v. 70, p. 024601, 2004.
%{Phys. Rev. C}, 70 (2004) 024601.
{Phys. Rev. C} {\bf 70}, 024601 (2004)

\bibitem{MERZBACHER} E. Merzacher, 
{\it Quantum Mechanics} (John Wiley, New York, 1961) %:621p, 1961.

\bibitem{PERES} A. Peres, 
%Nonexponential decay law.
%{\it Ann. Phys.} v. 129, n. 1, p. 33--46, 1980.
%{Ann. Phys.} 129(1) (1980) 33.
{Ann. Phys.} {\bf 129}(1), 33 (1980)

\bibitem{AVIGNONE} F. T. Avignone, 
%F. T.; Comment on
%'Test of the experimental decay law at short and long times'.
%{\it Phys. Rev. Lett.}   v. 61, n. 22, p. 2264, 1988.
%{Phys. Rev. Lett.} 61(22) (1988) 2264.
{Phys. Rev. Lett.} {\bf 61}(22),  2264 (1988)


\bibitem{GODOVIKOV} S. K. Godovikov,  
%Nonexponential $^{\rm 125m}$Te radiactive decay. {\it JETP Letters}
% v. 79, n. 5, p. 249--253, 2004.
%{JETP Lett.} 79(5) (2004) 249.
{JETP Lett.} {\bf 79}(5), 249 (2004)

\bibitem{SKOROBOGATOV} G. A. Skorobogatov, V. V. Eremin,  
%On the paper "Nonexponential $^{\rm 125m}$Te radiactive decay". {\it JETP Letters}
% v. 83, n. 1, p. 46--48, 2006.
%{JETP Lett.} 83(1) (2006) 46.
{JETP Lett.} {\bf 83}(1), 46 (2006)

\bibitem{ASTONa} P. J. Aston,  
%Is radiactive decay really exponential? 
%{\it EPL} v. 97, p. 52001, 2012.
%{EPL} 97 (2012) 52001.
{EPL} {\bf 97}, 52001 (2012)

\bibitem{ASTONb} P. J. Aston, 
%P. J. Reply to the comment by Cleanthes A. Nicolaides. 
%{\it EPL}  v. 101, p. 42002, 2013.
%{EPL} 101 (2013) 42002.
{EPL} {\bf 101}, 42002 (2013)

%\bibitem{NICOLAIDES} C. A. Nicolaides, %C. A. Comments on 'Is radiactive decay really exponential?'
%{\it EPL} v. 101, p. 42001, 2013.
%{\it EPL} 101 (2013) 42001.

\bibitem{ROTHEa} C. Rothe, A. P. Monkman, 
%Triplet exciton migration in a conjugated polyfluorene.
%{Phys. Rev. B}, v. 68, p. 075208/1--11, 2003.
%{Phys. Rev. B} 68 (2003) 075208.
{Phys. Rev. B} {\bf 68}, 075208 (2003) 

\bibitem{ROTHEb} C. Rothe, S. I. Hintschich, A. P. Monkman, 
%Violation of the exponential-decay law at long time.
%{\it Phys. Rev. Lett.}, v. 96, p. 163601/1--4, 2006.
%{Phys. Rev. Lett.} 96 (2006) 163601.
{Phys. Rev. Lett.} {\bf 96}, 163601 (2006) 

\bibitem{RIDA} S. Z. Rida, A. M. El-Sayed, A. A. M. Arafa, 
%Effect of bacterial memory dependent growth by using fractional derivatives reaction-diffusion
%chemotactic model.
%{J. Stat. Phys.}, v. 140, p. 797--811, 2010.
%{J. Stat. Phys.}, 140 (2010) 797.
{J. Stat. Phys.} {\bf 140}, 797 (2010)

\bibitem{POOSEH} {S. Pooseh, H. S. Rodrigues, D. F. M. Torres,} %Fractional derivatives in Dengue epidemics.}
%URL \url{arXiv:1108.1683v1 [math-CA] 8 Aug 2011}.
%NUMERICAL ANALYSIS AND APPLIED MATHEMATICS ICNAAM 2011: 
International Conference on Numerical Analysis and Applied Mathematics 
%AIP Conference Proceedings, 
%1389 (2011) 739.
{\bf 1389}, 739 (2011) 

\bibitem{DOKOUMETZIDIS} A. Dokoumetzidis, P. Macheras,  
%Fractional kinetics in drug absortion an disposition process.
% {\it J. Pahrmacokinet Pharmacodyn} v. 36, p. 165--178, 2009.
%{J. Pharmacokinet Pharmacodyn} 36 (2009) 165.
{J. Pharmacokinet Pharmacodyn} {\bf 36}, 165 (2009)


\bibitem{CALIK} A. E. Çalik, H. Ertik, B. \"Oder, H. Sirin,  
%A fractionl calculus approach to
%investigate the alpha decay process. 
%{\it Int. J. Mod. Phys. E}
%  v. 22, n. 7, p. 1350049, 2013.
{Int. J. Mod. Phys. E} {\bf 22}(7), 1350049 (2013) 

%\bibitem{MERZBACHER} E. Merzacher, {Quantum Mechanics}, John Wiley, New York, 1961. %:621p, 1961.
\bibitem{1965} K. S. Miller, B. Ross, {\it An Introduction to the Fractional Calculus and Fractional Differential Equations} 
(John Wiley \& Sons Inc., 
New York, 1993)

\bibitem{PODLUBNYbook} I. Podlubny,  {\it Fractional Direrential Equations, Mathematics on Science and Engeneering}, 
v. 198 (Academic Press, San Diego, 1999)

\bibitem{podart} {I. Podlubny, The Laplace Transform Method for Linear Differential Equations of the Fractional Order. }
%In modified form included in Chapters 4 and 5 of the book: Podlubny, I.: Fractional Differential Equations.    
{eprint arXiv:funct-an/9710005v1. }
%The Laplace Transform Method for Linear Differential Equations of the Fractional Order

\bibitem{abel} R. Gorenflo, S. Vessella, {\it Abel Integral Equations: Analysis and Applications} (Springer-Verlag, Berlin, 1991)

%\bibitem{ill} problema inverso?

\bibitem{Mainardi} {F. Mainardi, On some properties of the Mittag-Leffler function $E_\alpha(-t^\alpha)$, completely monotone for $t > 0$ with $0 < \alpha< 1$. eprint arXiv:1305.0161v3 [math-ph].}

%\bibitem{Mainardi} F. Mainardi, R. Gorenflo, 
%On Mittag-Leffer-type functions in fractional evolution processes. 
%{J. Comput. Appl. Math.}, 118 (2000) 283.

\bibitem{num} R. H. Landau, {\it Computational Physics Problem Solving with Computers} (John Wiley \& Sons Inc., 
New York, 1997)

\bibitem{podmat}{I. Podlubny, M. Kacenak, The Matlab mlf code. MATLAB CentralFile Exchange (2009). File ID: 8738.}

\bibitem{artigonovo} R. Garrappa, M. Popolizio, {Adv. Comput. Math.} {\bf 39} 205 (2013)


%%%%%%%%%%%%%%%%%%%%%%%%%%%%%%%%%%%%%%%%%%5
%\bibitem{CASASANTA} Casasanta, G.; Ciani, D.; Garra, R. Non-exponential extinction of radiation
%by fractional calculus modelling. {\it Journal of Quantitative Spectroscopy \& Radiative Transfer}
%  v. 113, p. 194, 2012.

%\bibitem{GAMOW} Gamow, G. Zur quantentheorie des atomkernes. {\it Z. Phys.} v. 51, n. 3(4), p. 204--212, 1928.

%\bibitem{MAINARDI} Mainardi, F.  On some properties of the Mittag-Leffer function $E_\alpha(-t^\alpha)$ completely monotone
%for $t>0$ with $0<\alpha<1$. %URL \url{arXiv:1305.0161v1 [math-ph] 1 May 2013}.

%\bibitem{MITTAG} Mittag

%\bibitem{NORMAN} Norman, E. B.; Gazes, S. B.; Crane, S. G. ; Bennett, D. A.
%Test of the experimental decay law at short and long times.
%{\it Phys. Rev. Lett.}   v. 60, n. 22, p. 2246--2249, 1988.

%\bibitem{NOVKOVIC} Novkovi\'c, D.; Nadded, L.; Kandi\'c, A.; Vukanac, I.; Durasevi\'c, M.; Jordanov, D.
%Testing the exponential decay law of gold $^{\rm 198}$Au.
%{\it Nucl. Instr. and Meth. A} v. 446, p. 477--480, 2006.

%\bibitem{PODLUBNYmatlab} Podlubny, I., MATLAB routine for evaluating the Mittag-Leffler function with two parameters. 
%URL \url{http://www.mathworks.com/matlabcentral/ fileexchange/8738-mittag-leffler-function}

%\bibitem{RUTHERFORD} Rutherford, E. Uranium radiation and the electrical conduction produced by
%it. {\it Phil. Mag.}, v. 47, n. 284, p. 109--163, 1899.

%\bibitem{SANTOS} dos Santos, J. P. C.; Cardoso, A.; Ferreira, E. C.; Franco, J. C.; SOUZA Jr., J. C. Cálculo de Ordem Fracionária e Aplicações.
%{\it Sigmae}, v.1, n.1, p. 18--32, 2012.

%\bibitem{SIMPLEX} Simplex

%\bibitem{SOBOL} Sobol, I. M. {\it A primer for the Monte Carlo}. London: CRC Press, 107p, 1994.

\end{thebibliography}
\end{document}